\documentclass[aps,prl,showpacs,amsmath,amssymb,lengthcheck]{revtex4}

\usepackage{epsfig}

\begin{document}

\title{Weakly Interacting Bose Mixtures at Finite Temperature}

\author{Bert Van Schaeybroeck}
\affiliation{Instituut
 voor Theoretische Fysica,\\ Katholieke Universiteit
Leuven,
 Celestijnenlaan 200 D, B-3001 Leuven, Belgium.}
\date{\small\it \today}

\begin{abstract}
Motivated by the recent experiments on Bose-Einstein mixtures with
tunable interactions we study repulsive weakly interacting Bose
mixtures at finite temperature. We obtain phase diagrams using
Hartree-Fock theory which are directly applicable to
experimentally trapped systems. Almost all features of the
diagrams can be characterized using simple physical insights. Our
work reveals two surprising effects which are dissimilar to a
system at zero temperature. First of all, no pure phases exist,
that is, at each point in the trap, particles of both species are
always present. Second, even for very weak interspecies repulsion
when full mixing is expected, \textit{condensate} particles of
both species may be present in a trap without them being mixed.
\end{abstract}

\pacs{03.75.Hh, 03.75.Mn, 67.60.Bc,67.85.Fg}

\maketitle

\textit{Introduction --} Since the early realization of mixtures
of Bose-Einstein condensates (BECs) ten years ago~\cite{stenger},
a host of experiments revealed the rich physics of two-component
Bose~\cite{pethick,pita} and Fermi~\cite{giorgini} systems.
Whereas most experiments use mixtures of hyperfine states of like
atomic species, it is also possible for unlike species to be
trapped simultaneously. Moreover both systems allow the use of
Feshbach resonances by means of which the interatomic interactions
can be tuned to arbitrary values and which revolutionized the
physics of ultracold gases. Such resonances were observed for
binary Bose gases consisting of
$^{87}$Rb$-^{41}$K~\cite{thalhammer},
$^{87}$Rb-$^{133}$Cs~\cite{pilch} and
$^{85}$Rb-$^{87}$Rb~\cite{papp,papp2} mixtures; clear phase
segregation was achieved for the last mixture by changing the
intraspecies interactions of the $^{85}$Rb particles~\cite{papp2}.

To date theoretical works on BEC mixtures predominantly focussed
on static and dynamic properties at zero temperature~\cite{ho}.
Interesting questions, however, can be posed concerning the finite
temperature regime: for instance, will, upon increasing the
interspecies repulsion, phase segregation for thermally depleted
particles set in at the same time as for condensate particles?
Does increasing the temperature induce an increased tendency to
mix the species? Can pure phases still exist? The issue of phase
segregation at nonzero temperature was already addressed in
Ref.~\onlinecite{shi} where the authors calculated the conditions
for an instability to occur in a volume with fixed particle
numbers. Also the case of a Bose mixtures in a trap at nonzero
temperature was briefly discussed.

In this Letter, we aim at clarifying the possible phase structures
in traps of Bose mixtures at finite temperature. The presented
phase diagrams are directly applicable to trapped situations, and
yet, independent of the specific trapping parameters. We realize
this by drawing the diagrams as a function of the chemical
potentials of the two species while taking the temperature to be
fixed. These diagrams are apt for gases confined by a smoothly
varying potential for then the use of a local density
approximation is appropriate and an effective chemical potential
can be taken as locally constant at each position in the trap.
Rather than giving exact diagrams using specific
parameters~\footnote{Our diagrams differ for different values of
(the later introduced) six parameters: $m_{_1}$, $m_{_2}$,
$a_{_{11}}$, $a_{_{22}}$, $a_{_{12}}$ and $T$. Since even the
scattering lengths vary in experiments, we limit our discussion to
generic features.}, we focus here on their basic understanding by
highlighting the generic features; more elaborate work will be
presented in a forthcoming paper. We argue that our diagrams can
be almost fully characterized, based on the knowledge of
single-species Bose systems at finite temperature and the BEC
mixture at zero temperature.

\textit{Equation of state} --- Consider two Bose species, labelled
$1$ and $2$, at temperature $T$ and at chemical potentials
$\mu_{_1}$ and $\mu_{_2}$. The particles of species $i$ and $j$
interact weakly via s-wave scattering, quantified by a positive
scattering length $a_{_{ij}}$ and a coupling constant
$G_{_{ij}}=2\pi\hslash^2a_{_{ij}}(m_i^{-1}+m_j^{-1})$ (henceforth
$i,j=1,2$). We use the self-consistent Hartree-Fock (HF) equations
of state as first introduced by Huang et al.~\cite{huang}. This
very model is known to give good agreement with
experiments~\cite{gerbier} and even with quantum Monte Carlo
simulations~\cite{holzmann}. The HF model treats noncondensed
particles as free particles with a mean field chemical potential
shift, and takes only thermal and no quantum fluctuations into
account. The HF grand potential per unit volume $\Omega$ is
expressed in terms of the condensate densities $n_{_{ci}}$ and
thermally depleted densities $n_{_{di}}$~\cite{huang, pita}:
\begin{widetext}
\begin{align}%\label{fase_mixedenergy}
\Omega=\sum_{i=1,\,2}\left[-\frac{g_{_{5/2}}\left(e^{\beta\mu_{_i}^{_0}}\right)
k_{_B}T}{\lambda_{_{i}}^3}+\mu_{_i}^{_0}n_{_{di}}+G_{_{ii}}\left(\frac{n_{_{ci}}^2}{2}
+2n_{_{ci}}n_{_{di}}+n_{_{di}}^2\right)-\mu_{_i}(n_{_{ci}}+n_{_{di}})\right]
+G_{_{12}}(n_ {_{c1}}+n_{_{d1}})(n_ {_{c2}}+n_{_{d2}}),\nonumber
\end{align}
\end{widetext}
where $\lambda_{_{i}}=\sqrt{2\pi\hslash^2/(m_{_i} k_{_B}T)}$,
$\beta=1/k_{_B}T$ and $g_{_l}(x)=\sum_{_{s=1}}^{_\infty} s^{-l}
x^l$. The first two terms are due to the entropy and the kinetic
energy of the depleted particles. The very last term couples the
two species by repulsive interactions. At fixed chemical
potential, minimization of $\Omega$ with respect to the densities
$n_{_{ci}}$ and $n_{_{di}}$ yields the self-consistent
Hartree-Fock equations of state:
\begin{subequations}\label{EL1}
\begin{align}
\mu_{_1}&=\mu_{_1}^{_0}+2G_{_{11}}(n_{_{c1}}+n_{_{d1}})+G_{_{12}}(n_{_{c2}}+n_{_{d2}}),\\
\mu_{_2}&=\mu_{_2}^{_0}+2G_{_{22}}(n_{_{c2}}+n_{_{d2}})+G_{_{12}}(n_{_{c1}}+n_{_{d1}}),
\end{align}
\end{subequations}
and, $\mu_{_i}^{_0}=-G_{_{ii}}n_{_{ci}}$ whenever $n_{_{ci}}\neq
0$. The depleted densities on the other hand are governed by
minimization with respect to $\mu_{_i}^{_0}$:
\begin{align}\label{EL2}
n_{_{di}}=g_{_{3/2}}(e^{\beta\mu_{_i}^{_0}})/\lambda_{_{i}}^3.
\end{align}

For most systems of ultracold gases only few lengthscales are
relevant. These include the particle de Broglie wavelength
$\lambda_{_{i}}$, the interparticle distance $n_{_i}^{-1/3}$ and
lastly the scattering lengths $a_{_{ij}}$. The HF theory as
introduced above is valid in case of weak interactions i.e. for
small values of $a_{_{ij}}/\lambda_{_i}$ and when
\textit{single}-particle excitations prevail i.e. when $n_{_i}
a_{_{ij}}\lambda_{_{i}}^2\ll 1$~\cite{pethick}. Also, the theory
fails very close to the transition point where fluctuations become
dominant. In nowadays experiments, the scattering length takes
values of order of ten nanometers and below, whereas the
interparticle distances and the Broglie wavelengths are typically
of the order of microns and higher. It follows that the necessary
conditions for the HF theory to be valid are
fulfilled~\footnote{The scattering length is known to diverge near
a Feshbach resonance. Nevertheless, our theory still applies to
the experiments of Ref.~\onlinecite{papp2} since the largest
values of $n_{_i} a_{_{ij}}\lambda_{_{i}}^2$ are around $0.1$.}.

We proceed by first discussing BEC mixtures at zero temperature
and a single species Bose gas at finite temperature, since these
systems provide all ingredients necessary for the understanding of
the results obtained further.

\textit{BEC mixture at $T=0$ ---} At zero temperature, the HF
theory reduces to Gross-Pitaevskii (GP) theory for a spatially
homogeneous system. According to GP theory, two regimes can be
distinguished, depending on the parameter $\Delta$ defined
as~\cite{ho}:
\begin{align}
\Delta= G_{_{11}}G_{_{22}}/G_{_{12}}^2-1.
\end{align}
When $\Delta<0$, the inter-species repulsion is sufficiently
strong to induces phase segregation and solely BEC $1$ or BEC $2$
may exist as ground states, that is, species $1$ and $2$ are
immiscible. One readily finds the following conditions:
\begin{subequations}\label{lines1}
\begin{align}
&\text{BEC 1 when } \mu_{_1}>\mu_{_2}\sqrt{G_{_{11}}/G_{_{22}}},\\
&\text{BEC 2 when } \mu_{_1}<\mu_{_2}\sqrt{G_{_{11}}/G_{_{22}}}.
\end{align}
\end{subequations}
On the other hand, when $\Delta>0$, the interspecies repulsion is
weak and a mixed phase of nonzero density of BEC $1$ and BEC $2$,
here denoted as BEC MIX, may appear as the ground state:
\begin{align}\label{lines2}
\text{BEC MIX when }
G_{_{12}}/G_{_{22}}<\mu_{_1}/\mu_{_2}<G_{_{11}}/G_{_{12}}.
\end{align}

\textit{Single-species at finite $T$ ---} Consider now a gas of
species $i$ at fixed temperature. One can think of the chemical
potential $\mu_{_i}$ as a measure for the particle density: the
higher $\mu_{_i}$, the higher the density of species $i$. For
large and negative values of $\mu_{_i}$, the system is dilute and
the interaction energy small. On the other hand, above a certain
``critical'' positive chemical potential a transition to a phase
with nonzero condensate density sets in; by the HF approach, this
critical value is attained when $\mu_{_i}^{_0}=0$ such
that~\cite{pita}~\footnote{Small corrections apply near the
critical points, due to the first-order nature of the transition;
as mentioned before, the theory is not accurate very close to the
transition point.}:
\begin{align}\label{critchem}
\mu_{_i}^{_C}=2G_{_{ii}}\,\zeta(3/2)/\lambda_{_i}^3,
\end{align}
with $\zeta$ the Riemann zeta function. Note that $\mu_{i}^{_C}$
depends on temperature via $\lambda_{_i}$ and from the scaling
$\mu_{_i}^{_C} \simeq k_{_B} T(a_{_{ii}}/\lambda_{_i})$ we
conclude that $\mu_{_i}^{_C}\ll k_{_B} T$.

\begin{figure}
  \epsfig{figure=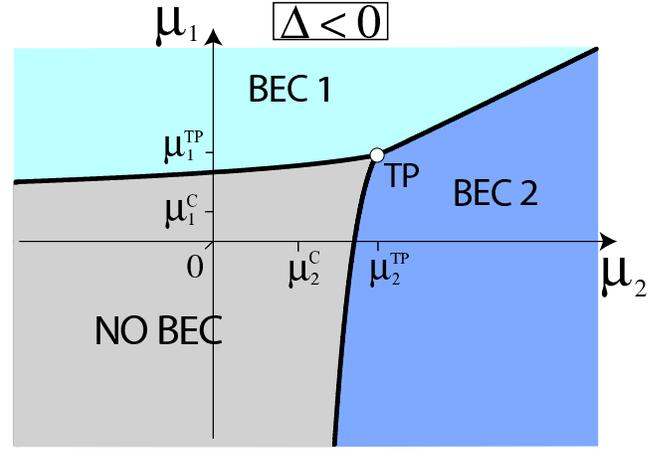,angle=0, width=240pt}
  \caption{\label{fig1} (Color online) Generic phase diagram for two Bose species at fixed temperature in case of strong interspecies repulsion
  ($\Delta<0$) as a function of the chemical potentials of both species. The BEC 1 (green) phase consists of condensate particles of
  species 1 and thermally depleted particles of species 1 and 2 and vice versa for the BEC 2 (blue) phase. The (grey) phase NO BEC consists of depleted
  particles of both species. The
  chemical potential $\mu_{_i}^{_C}$ is given by Eq.~\eqref{critchem} and the locus of the triple point TP is given in Eq.~\eqref{triple}
  while the first-order phase boundaries are described by Eqs.~\eqref{lines3},~\eqref{lines4} and~\eqref{lines5}.
  }
\end{figure}

The introduction of a second Bose gas will shift the critical
chemical potential to a higher value than $\mu_{_i}^{_C}$; this
can be ascribed to the interspecies repulsion which tends to
rarify the gas of species $i$.

\begin{figure}
  \epsfig{figure=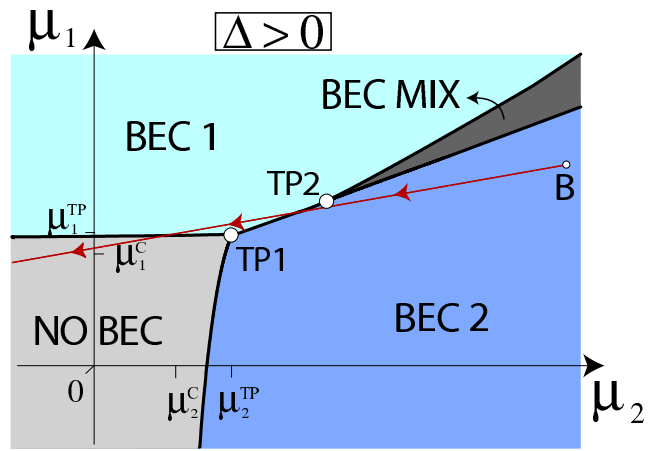,angle=0, width=240pt} \caption{(Color online) The same
applies as in Fig.~\ref{fig1} except that weak interspecies
interactions are considered ($\Delta>0$). The (dark grey) BEC MIX
phase consists of condensed and depleted particles of both
species. The red arrowed line indicates a possible tracing of the
diagram as a function of the (radial) position in the trap. The
point B indicates the chemical potentials at the center of the
trap.\label{fig2}
  }
\end{figure}

\textit{Phase Diagram for $\Delta<0$ and finite $T$ ---} Upon
working at fixed particle numbers as in Ref.~\onlinecite{shi}, the
onset of phase segregation can be naturally probed by the presence
of a density instability. At fixed chemical potentials on the
other hand, this method is inadequate as an instability indicates
a spinodal (line) rather than the transition itself in case it is
of first order. Instead, we have compared numerically the grand
potentials for the eight possible thermodynamical phases and their
associated multiple solutions. We arrive at the generic phase
diagrams for Bose mixtures at finite temperature in
Figs.~\ref{fig1} and~\ref{fig2} for the cases $\Delta<0$ and
$\Delta>0$ respectively.

Identical to the behavior at $T=0$, no mixing of unlike condensate
particles is possible at finite $T$ when $\Delta<0$. Only three
phases appear in Fig.~\ref{fig1}, none of which are pure; that is,
even for extremely strong interspecies repulsion, depleted
particles of both species are present in all phases. Accordingly,
BEC $1$ denotes the phase composed of condensate particles of
species $1$ and depleted particles of both species, and vice versa
for BEC $2$. The NO BEC phase denotes the phase with thermal
particles of species $1$ and $2$.

A triple point TP is present in the phase diagram of
Fig.~\ref{fig1}. As checked numerically, its locus
$(\mu_{_1}^{_{TP}},\mu_{_2}^{_{TP}})$ can be found by setting
$\mu_{_1}^{_0}=\mu_{_2}^{_0}=0$ in Eq.~\eqref{EL1}:
\begin{subequations}\label{triple}
\begin{align}
\mu_{_1}^{_{TP}}&=\mu_{_1}^{_C}+\mu_{_2}^{_C}G_{_{12}}/2G_{_{22}},\\
\mu_{_2}^{_{TP}}&=\mu_{_2}^{_C}+\mu_{_1}^{_C}G_{_{12}}/2G_{_{11}}.
\end{align}
\end{subequations}
Note that in units of $\mu_{_1}^{_C}$ and $\mu_{_2}^{_C}$, the
position of point TP is temperature independent.

Consider now the phase boundary between BEC 1 and 2; for large and
positive values of both chemical potentials, the interaction
energy per particle will eventually become much larger than the
thermal energy and so asymptotically the $T=0$ behavior is
expected. Indeed, we find that in accord with Eq.~\eqref{lines1}
the BEC 1-2 phase boundary satisfies:
\begin{align}\label{lines3}
\mu_{_1}/\mu_{_2}=\sqrt{G_{_{11}}/G_{_{22}}} \text{ when both }
\mu_{_1},\mu_{_2}\rightarrow \infty.
\end{align}
Moving along the NO BEC-BEC 1 boundary away from TP, the density
of species $2$ decreases; in the limit of $\mu_{_2}\rightarrow
-\infty$, only particles of species $1$ remain such that the
single-component value $\mu_{_1}\rightarrow\mu_{_1}^{_C}$ is
attained. The associated asymptotic decay is exponential as the NO
BEC-BEC 1 phase boundary is described by:
\begin{align}\label{lines4}
\mu_{_1}=\mu_{_1}^{_C}+G_{_{12}}e^{\mu_{_2}/k_{_B}T}/\lambda_{_2}^3\text{
when } \mu_{_2}\rightarrow -\infty.
\end{align}
Because $\mu_{_2}^{_C}\ll k_{_B} T$, it follows that the
asymptotic convergence is slow if $\mu_{_2}$ is expressed in units
of $\mu_{_2}^{_C}$. The same line of argument with interchanged
indices $1$ and $2$ applies to the demarcation line between BEC 2
and NO BEC:
\begin{align}\label{lines5}
\mu_{_2}=\mu_{_2}^{_C}+G_{_{12}}e^{\mu_{_1}/k_{_B}T}/\lambda_{_1}^3\text{
when } \mu_{_1}\rightarrow -\infty.
\end{align}

In outline, we have shown that one can qualitatively establish the
phase diagram for the case $\Delta<0$ without explicitly
performing any numerics. Indeed, we found that the expressions for
the triple point (Eq.~\eqref{triple}), and the phase boundaries
(Eqs.~\eqref{lines3}-\eqref{lines5}) agree well with the
numerically exact results.

\textit{Phase Diagram for $\Delta>0$ at finite $T$---} At zero
temperature mixing of unlike condensate particles is possible when
$\Delta$ is positive. We find the same behavior at finite $T$.
This is shown in Fig.~\ref{fig2} where BEC MIX consists of
condensed and depleted particles of both species. The bifurcation
of the BEC 1-2 line brings along the appearance of a new triple
point TP2 (TP1 is identical to TP of Fig.~\ref{fig1}). For the
aforementioned reasons, the $T=0$ physics is recovered for large
and positive chemical potentials. Hence Eq.~\eqref{lines2} implies
that the BEC MIX phase is delimited by the phase boundaries
\begin{align}\label{lines6}
\begin{cases}
\mu_{_1}/\mu_{_2}=G_{_{11}}/G_{_{12}}\\
\mu_{_1}/\mu_{_2}=G_{_{12}}/G_{_{22}}
\end{cases}
 \text{when both }
\mu_{_1},\mu_{_2}\rightarrow\infty.
\end{align}

The location of triple point TP2 alters upon varying temperature
and $\Delta$. For small and positive values of $\Delta$, TP2
enters at large and positive values of the chemical potentials
while TP2 coalesces with TP1 only in the limit
$\Delta\rightarrow\infty$. This leads us to a remarkable
conclusion: even when the zero temperature condition for species
mixing is satisfied, at finite temperature, two-phase equilibrium
between BEC 1 and 2 is still allowed. Changing the temperature
also induces a displacement of TP2 with respect to TP1.
Intuitively one expects that increasing the temperature favors
mixing. Yet, condensate particles carry no entropy and we find
rather the opposite: increasing the temperature suppresses mixing.
In particular, the distance between TP1 and TP2 (in units of
$\mu_{_i}^{_C}$) increases upon increasing the temperature.

In the ideal case of fully suppressed interspecies interactions
($\Delta\rightarrow\infty$) the triple points TP1 and TP2 coincide
and, as expected, the phase boundaries are the horizontal and
vertical lines $\mu_{_1}=\mu_{_1}^{_C}$ and
$\mu_{_2}=\mu_{_2}^{_C}$.

%~\footnote{Note also that, whereas the phase boundaries associated
%with Eqs.~\eqref{lines1} and \eqref{lines2} are of first and
%second order respectively, they are all of first order at finite
%temperature.}.

To summarize, the $\Delta>0$ phase diagram is similar to the
$\Delta<0$ diagrams but is marked by the appearance of an
additional phase consisting of mixed BEC species, and an
additional triple point TP2. Despite the known asymptotic behavior
of the phase boundaries around the BEC MIX phase (see
Eq.~\eqref{lines5}), the locus of TP2 depends in a nontrivial way
on temperature and $\Delta$ and must therefore be determined
numerically.

\textit{Discussion ---} In order to extract from Figs.~\ref{fig1}
and~\ref{fig2} the possible phase structures appearing in an
experimental trap, the trapping potentials $U_{_i}$ are required.
In most experiments a local density approximation or Thomas-Fermi
approximation is justified such that at each position $\mathbf{r}$
an effective chemical potential
$\mu_{_i}(\mathbf{r})=\mu_{_i}-U_{_i}(\mathbf{r})$ may be assumed
constant; $\mu_{_i}(\mathbf{r})$ is maximal at the center and
minimal at the edge of a trap. In case both species are
harmonically confined or
$U_{_i}(\mathbf{r})=m_{_i}\omega_{_i}^2\mathbf{r}^2/2$, the
chemical potentials probe our phase diagrams following a straight
line:
\begin{align}
m_{_2}\omega_{_2}^2(\mu_{_1}(\mathbf{r})-\mu_{_1})=m_{_1}\omega_{_1}^2(\mu_{_2}(\mathbf{r})-\mu_{_2}).
\end{align}
As an example, a trap configuration which is described by the red
path in Fig.~\ref{fig2} contains a BEC 2 core surrounded by shells
of BEC 1 and NO BEC. Note that our phase diagrams do not
explicitly depend on trapping parameters nor gravity.

%~\footnote{Note that in case of harmonic confinement, the local
%density approximation is justified when
%$\lambda_{_i}\ll\sqrt{\hslash/m_{_i}\omega_{_i}}$.}.

The most striking conclusion of this work is that no pure phases
exist at finite temperature; even for large interspecies
interactions, the depleted particles are not entirely
expelled~\footnote{This very effect therefore vanishes at zero
temperature when the depleted densities are quenched.}. However,
the fact that upon increasing the interspecies repulsion
$G_{_{12}}$ in the BEC MIX phase, condensate particles will be
expelled first, may be understood from the following simplistic
argument: in the expression for $\Omega$, it is seen that due to
different exchange terms the intraspecies interactions for
depleted particles are a factor $2$ higher than those of
condensate particles. A naive application of the zero-temperature
criterion for phase segregation
($G_{_{12}}>\sqrt{G_{_{11}}G_{_{22}}}$) to depleted particles,
yields segregation for values of $G_{_{12}}$ higher than
$\sqrt{2G_{_{11}}G_{_{22}}}$.

As a direct extension of this work, the interface physics and its
impact on the phase structures can be explored in case of
two-phase equilibrium. A generalization of our HF theory to
spatially inhomogeneous systems is thereby required; within a
first approximation, this can be effectuated by introducing the
terms $\boldsymbol{\nabla}^2\sqrt{n_{_{ci}}}$ into the grand
potential~\cite{pita}. Such theory would then allow the
calculation of surface excess quantities which are useful for
tightly confined gases where their effect on the ground state
configuration turns out to be
substantial~\cite{vanschaeybroeck,papp2,partridge}. Further
challenges include a generalization of the theory of anomalous
wetting phase transitions to finite temperatures~\cite{indekeu}.

%Note finally that our theory is also applicable for like atomic
%species with two different hyperfine states or after converting a
%large fraction of atoms into a molecular state~\cite{donley}.

%The absence of pure phases may be an artifact of the HF model,
%however, we expect that the phase diagrams are qualitatively very
%reliable. HF model is not valid very close to critical points
%where fluctuations become important. Within limits of HF

\textit{Conclusion} --- We have elaborated generic phase diagrams
for Bose mixtures at finite temperature based on the Hartree-Fock
model. As corroborated by numerical analysis, almost all
properties of our phase diagrams can be expressed in terms of
known results for single-species gases at finite temperature and
for binary mixtures at zero temperature. This means that it is
possible to establish the phase diagrams (aside from the triple
point TP2 in case $\Delta>0$) without performing any numerics. The
phase diagrams are drawn as a function of the chemical potentials
of the species such that, spatially tracing the trap is equivalent
to exploring the phase diagram. We highlight the importance of the
triple points in our diagrams and find that, surprisingly, no pure
phases exist, and, increasing the temperature tends to suppress
the mixing of condensate particles.

%Increasing the temperature has the effect of
% eventueel: no real phase segregation but rather BEC transition

%In particular,

\textit{Acknowledgement} --- The author thanks Lev Pitaevskii,
Joseph Indekeu and Achilleas Lazarides for useful suggestions.
B.V.S. is a Postdoctoral Fellow of FWO.

\end{document}